\documentclass[conference]{IEEEtran}
\IEEEoverridecommandlockouts
\usepackage{cite}
\usepackage{amsmath,amssymb,amsfonts}
\usepackage{algorithmic}
\usepackage{graphicx}
\usepackage{textcomp}
\usepackage{xcolor}
\usepackage{array}
\usepackage{subfloat}
\usepackage{subfigure}
\newtheorem{definition}{Definition}

\def\BibTeX{{\rm B\kern-.05em{\sc i\kern-.025em b}\kern-.08em
    T\kern-.1667em\lower.7ex\hbox{E}\kern-.125emX}}

\begin{document}
\title{Age-Energy Trade-off in Status Update System with Wake-up Control}
% \author{Jiajie Huang and Jie Gong, Member, IEEE}%
\author{\IEEEauthorblockN{Jiajie Huang, Jie Gong}
\IEEEauthorblockA{\textit{Guangdong Key Laboratory of Information Security Technology,} \\
\textit{School of Computer Science and Engineering, Sun Yat-sen University}, Guangzhou, 510006, China \\
Email: huangjj7@mail2.sysu.edu.cn, gongj26@mail.sysu.edu.cn}
\thanks{This research was funded by the National Natural Science Foundation of China grant number 62171481, the National Key Research and Development Program of China grant number 2019YFE0114000, the Special Support Program of Guangdong grant number 2019TQ05X150, the Natural Science Foundation of Guangdong Province grant number 2021A1515011124, and the Science and Technology Program of Guangzhou under Grant 202201011577.}
}

\maketitle

\begin{abstract}%
In the status update system, the freshness of information is drawing more and more attention. To measure the freshness of the data, age-of-synchronization (AoS) is introduced. Since many communication devices are energy-constrained, how to reduce energy consumption while keeping the timely response of data needs to be carefully addressed. In this paper, we study the age-energy trade-off problem in a single-source single-server scenario. We assume the server enters a low-power sleep state when idle to save energy and consider three wake-up policies. We adopt the stochastic hybrid system (SHS) method to analyze the average AoS and average energy consumption under different policies. The age-energy trade-off relationship under different parameters is illustrated by numerical results.
\end{abstract}
%p

\section{Introduction}\label{0}
In the status update system, it is essential to know the state changes of the source in time. As a result, many indicators to measure the freshness of information have been proposed. Conventionally, age-of-information (AoI)\cite{AOI} has been widely studied, which is defined as the time elapsed since the latest successfully accepted update was generated. Recently, age-of-synchronization (AoS) is proposed to track whether the data is synchronized, which measures the time elapsed since latest information at the receiver becomes desynchronized \cite{AOS}. Compared with the AoI metric, AoS is more suitable for scenarios where the monitored data is updated less frequently, such as databases, web crawling systems and error alarm systems. For the differences in application scenarios between AoI and AoS, many performance studies on AoI are not very suitable for AoS. Therefore, it is meaningful to study how to obtain good AoS performance.

At the same time, due to the deployment of more and more communication equipment, energy consumption has become one of the critical issues for information and communication technology (ICT). Besides improving the freshness of information, reducing the cost of communication energy consumption also arouses widespread concern. However, improving information freshness and saving energy consumption usually conflict with each other. Intuitively, to keep the data fresh, it is necessary to process it in time, which means that the equipment consumes more energy. Therefore, there is an age-energy trade-off problem, and studying this problem has important guiding significance in the state update system with limited energy consumption.

In recent years, there have been several studies on the trade-off between information freshness and energy consumption. Energy harvesting sources were considered in Ref. \cite{energy1}, while B. T. Bacinoglu et al. studied age optimal strategies under infinite battery \cite{energy2}, unit battery \cite{energy3}, and finite battery \cite{energy4}, respectively. The trade-off between energy consumption and AoI has recently gained more attention \cite{energy11,energy12}, and the age-energy trade-off has been studied in error-prone channels \cite{gj} and in fading channels \cite{energy9}. Ref. \cite{hjj} analyzed the age-energy trade-off in a state update system based on hybrid automatic repeat request (HARQ). However, to the best of our knowledge, the trade-off between AoS and energy consumption is an open problem and has not been studied yet.

The SHS method is an effective and simple approach for analyzing data freshness over networks, as proposed in \cite{SHS for AoI}. It has been applied to various scenarios and queuing models. For instance, \cite{SHS1} studies the average AoI of each node in a single-source multi-hop status update system. The authors of \cite{SHS2} consider a multi-source FCFS M/M/1 queuing model with an infinite queue length. In \cite{SHS3}, a multi-server and multi-source LCFS queuing model is considered, with preemption in service adopted. The authors of \cite{SHS4} consider a status update system with two sources and propose three packet management strategies. The average AoI under each strategy is deduced by the SHS method. However, energy consumption has not been taken into account in any of these works. Therefore, this paper proposes to use the SHS method to analyze energy consumption and AoS simultaneously.

In this paper, we focus on the trade-off between AoS and energy consumption. We first introduce the system model and the definition of AoS. Secondly, a sleep model is introduced and three wake-up policies are proposed. Then, the SHS method is briefly introduced, and the average AoS and average energy consumption of the three wake-up policies are analyzed by SHS. Finally, numerical simulation shows the age-energy trade-off of different policies.

\section{System Model}\label{I}%
\begin{figure}[htb]
\centering
\includegraphics[width=80mm]{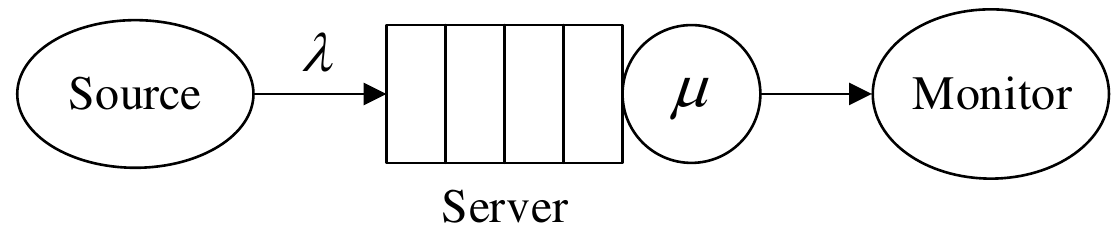}
\caption{System model.}
\label{system model}
\end{figure}

In this work, we consider a single-source single-server real-time status update system, as shown in Fig. \ref{system model}. In this system, the status updates of the source are generated randomly, which follows Poisson process with parameter $\lambda$. The updates are sent to the server in the form of data packets. The server processes the data packets at an exponentially distributed service rate with parameter $\mu$, and sends them to the monitor. The preemption strategy is adopted, that is, any newly generated packets from the source directly preempt the ones being processed by the server.

\begin{figure}[htb]
\centering
\includegraphics[width=60mm]{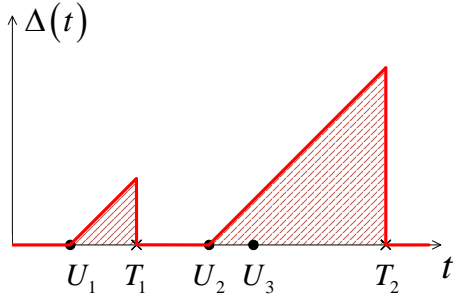}
\caption{Sample AoS path. $ \bullet $ indicates the state update at the source and $ \times $ indicates the state update at the monitor is synchronized with the source.}
\label{AoS path}
\end{figure}

We adopt the AoS \cite{AOS} as the indicator of the freshness of the state information of the physical process from the source. Fig. \ref{AoS path} describes a sample path for source updates and monitor synchronization. Let's define ${U_1},{U_2}, \ldots {U_k}$ as the sequence of source update times, ${T_1},{T_2}, \ldots {T_k}$ as the sequence of monitor refresh times. In addition, we denote $N(t)$ as the number of source refreshes up to time $t$. Formally, the definition of AoS is as follows.
\begin{definition}
    \emph{Let ${u}\left( t \right)$ denote the earliest time that the source gets a state update since the last refresh of the monitor copy, i.e.,
    \begin{equation}
        {u}\left( t \right) = \min \left\{ {\left. {{U_k}} \right|{U_k} > {T_{N\left( t \right)}}} \right\}.
        \label{eq01}
    \end{equation}
    The AoS at time t is defined as
    \begin{equation}
        \Delta(t) = \max (t - u(t),0).
        \label{eq1}
    \end{equation}
    Note that if the monitor's update is the same as the source, then $\Delta(t) = 0$.}
\end{definition}

It can be seen from (\ref{eq1}) that whenever a new status update is generated from the source, the monitor becomes unsynchronized with the source and the AoS of the source starts to increase. The AoS value drops to zero when a new packet is processed and received by the monitor, and remains at zero until a new status update is generated. In our work, we adopt the average AoS of the source over time, denoted by $\bar \Delta = {\lim _{t \to \infty }}E\left[ {\Delta \left( t \right)} \right]$, as a metric to evaluate the system performance.

\begin{figure}[htb]
\centering
\includegraphics[width=70mm]{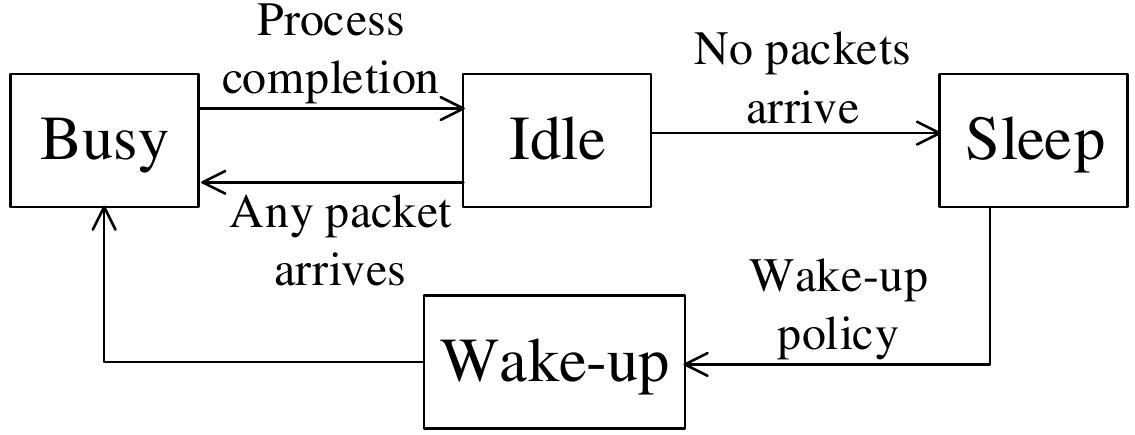}
\caption{Sleep model.}
\label{sleep model}
\end{figure}

The energy consumption of the server is another important performance metric. Therefore, we can adopt some sleep-wakeup policies to save energy when there is no updates to be processed. The sleep model is illustrated in Fig. \ref{sleep model}, where the server has four states: \emph{busy}, \emph{idle}, \emph{sleep} and \emph{wake-up}. In the busy state, the server is processing data packets. After all the packets in the server are processed, the server enters the idle state and operates at a low power level. If any new packets arrive during the idle state, the server immediately enters the busy state and starts processing them without delay or energy cost. If no new packets arrive during the entire idle state, the server enters the sleep state with extremely low power consumption, which can be considered as zero. We assume that the duration of the idle state follows an exponential distribution with mean $d$. The server transfers from the sleep state to the wake-up state based on some wake-up policies. After a certain amount of time in the wake-up state, the server turns to the busy state and begins to process the packet. The time cost of the wake-up state is assumed to follow an exponential distribution with mean $\theta$.

In our work, we consider the energy consumption rate ${\rm E}\left[ P \right]$ as another performance metric. Specifically, we assume that the energy per unit time consumed by the server in each state is denoted by ${P_{\textrm{B}}}$, ${P_{\textrm{I}}}$, ${P_{\textrm{S}}}$ and ${P_{\textrm{W}}}$ respectively. In general, ${P_{\textrm{B}}}$ is the largest, ${P_{\textrm{S}}}$ is the smallest, and ${P_{\textrm{I}}}$ and ${P_{\textrm{W}}}$ are between them.

\subsection{Wake-up Policy}
% The average energy consumption and the average AoS depend on the wake-up policy. In this work, we consider three wake-up policies as described below.
Different wake-up policies in queueing systems have been widely discussed in literatures, such as N-policy \cite{N-po}, single-sleep \cite{sin-sl}, multiple-sleep \cite{mul-sl}. In this work, we consider these three wake-up policies, which are described in detail below.

\begin{itemize}
\item[$\bullet$] \textit{\textbf{N-policy}}: Under this wake-up policy, the server remains in the sleep state until $N$ packets arrive. Once the ${N^{th}}$ packet arrives, it immediately transfers to the wake-up state.
\item[$\bullet$] \textit{\textbf{Single-sleep}}: Under this wake-up policy, the server turns to the wake-up state when it stays in the sleep state for a certain period of time. This period is assumed to follow an exponential distribution with mean $s$. It is worth noting that if there is no packet arrival during the wake-up state, the server does not immediately turn into busy, but keep idle until a packet arrives.
\item[$\bullet$] \textit{\textbf{Multi-sleep}}: This wake-up policy can be seen as an extension of the Single-sleep. In particular, the server firstly sleeps for a period of time after entering the sleep state. If no data packets arrive during this period of time, the server sleeps again for another period of time. The procedure repeats until some data packets arrive. Then, the server transfers to the wake-up state after the end of the current sleep period. Each time period is also assumed to follow an exponential distribution with mean $s$.
\end{itemize}

In this paper, we aim to analyze the AoS and energy performance of the above system with different wake-up policies. To study the problem with a unified framework, we adopt the SHS method for analysis, which is introduced in the next section.

\section{Analysis with Stochastic Hybrid System}
In this section, we first briefly introduce the SHS method and show how to use SHS to analyze the average AoS and the average energy consumption. Then, we show the analytical results with three wake-up policies respectively.

\subsection{SHS method}
\subsubsection{A Brief Introduction of SHS}
\
\newline
\indent SHS is a kind of stochastic dynamic system which combines continuous change with discrete state variation, and the change of system structure is related to some transformation rules\cite{SHS}. The evolution of the discrete state is determined by the transition or reset mapping, while the evolution of the continuous state is determined by the stochastic differential equation. The transition of discrete state is generally triggered by random events, and the probability of transition at a given time depends on the continuous and discrete components of the current SHS state. Therefore, SHS can be viewed as a piecewise deterministic Markov process in a certain sense. According to the definition of stochastic process, SHS can be expressed as
\begin{equation}
    \frac{{d{\mathbf{x}}\left( t \right)}}{{dt}} = f\left( {q\left( t \right),{\mathbf{x}}\left( t \right),t} \right) + g\left( {q\left( t \right),{\mathbf{x}}\left( t \right),t} \right)\frac{{d{\mathbf{z}}\left( t \right)}}{{dt}}.
    \label{eq2}
\end{equation}%
Where the discrete state is $q\left( t \right) \in \mathbb{Q}$, $\mathbb{Q}$ is a discrete set. The continuous state is ${\mathbf{x}}\left( t \right) \in {\mathbb{R}^{n + 1}}$. ${\mathbf{z}}\left( t \right)$ describes the process of independent Brownian motion. Thus, there is a mapping $f:\mathbb{Q} \times {\mathbb{R}^{n + 1}} \times \left[ {0,\infty } \right) \to {\mathbb{R}^{n + 1}}$ and $g:\mathbb{Q} \times {\mathbb{R}^{n + 1}} \times \left[ {0,\infty } \right) \to {\mathbb{R}^{\left( {n + 1} \right) \times k}}$. And there is a set of transitions $L$, each $l \in L$ defines a discrete transition/reset map ${\phi _l}:\mathbb{Q} \times {\mathbb{R}^{n + 1}} \times \left[ {0,\infty } \right) \to \mathbb{Q} \times {\mathbb{R}^{\left( {n + 1} \right) \times k}}$. Therefore, the state transition is
\begin{subequations}\label{eq3}
\begin{equation}
    \left( {q'\left( t \right),{\mathbf{x'}}\left( t \right)} \right) = {\phi _l}\left( {q\left( t \right),{\mathbf{x}}\left( t \right),t} \right).
    \label{eq3a}
\end{equation}%
The corresponding transition intensity is
\begin{equation}
    {\lambda ^{\left( l \right)}}\left( {q\left( t \right),{\mathbf{x}}\left( t \right),t} \right),\quad{\lambda ^{\left( l \right)}}:\mathbb{Q} \times {\mathbb{R}^{n + 1}} \times \left[ {0,\infty } \right) \to \left[ {0,\infty } \right).
    \label{eq3b}
\end{equation}
\end{subequations}

When the system is in a discrete state, the continuous state evolves according to (\ref{eq2}). When the discrete state of the system changes from $q$ to $q'$, the continuous state jumps from $\mathbf{x}$ to $\mathbf{x'}$ according to (\ref{eq3a}), and the frequency of the transition is determined by (\ref{eq3b}). In practice, the transition intensity is generally the instantaneous rate at which the transition occurs.
\\
\subsubsection{SHS for AoS and Energy Consumption}
\
\newline
\indent When using SHS to describe the AoS, the discrete state $q\left( t \right)$ represents the server occupancy, while the continuous state $\mathbf{x}\left( t \right)$ represents the deterministic constant slope ramp process. Therefore, for the general SHS model given in (\ref{eq2}) and (\ref{eq3}), we have
\begin{subequations}\label{eq4}
\begin{equation}
    f\left( {q\left( t \right),{\mathbf{x}}\left( t \right),t} \right) = {{\mathbf{b}}_q},
    \label{eq4a}
\end{equation}%
\begin{equation}
    g\left( {q\left( t \right),{\mathbf{x}}\left( t \right),t} \right) = 0,
    \label{eq4b}
\end{equation}%
\begin{equation}
    {\lambda ^{\left( l \right)}}\left( {q\left( t \right),{\mathbf{x}}\left( t \right),t} \right) = {\lambda ^{\left( l \right)}}{\delta _{{q_l},q}},
    \label{eq4c}
\end{equation}%
\begin{equation}
    {\phi _l}\left( {q\left( t \right),{\mathbf{x}}\left( t \right),t} \right) = \left( {{{q'}_l}\left( t \right),{\mathbf{x}}\left( t \right){{\mathbf{A}}_l}} \right).
    \label{eq4d}
\end{equation}%
\end{subequations}

In the Markov chain $q\left( t \right)$, each state $q \in \mathbb{Q}$ is a node on the chain, and the transition between states $l$ is a directed edge $({q_l},{q'_l})$ with a transition rate of ${\lambda ^{\left( l \right)}}$. The Cronecker function $\delta$ in (\ref{eq4c}) guarantees that the transition $l$ occurs only in the state $q_l$. When a state transition occurs, the discrete state $q_l$ changes to the state $q'_l$, and the continuous state ${\mathbf{x}}\left( t \right)$ is transformed according to the binary transfer reset mapping matrix ${{\mathbf{A}}_l}$: ${\mathbf{x'}}\left( t \right)={\mathbf{x}}\left( t \right){{\mathbf{A}}_l}$. In addition, according to (\ref{eq4a}) and (\ref{eq4b}), the evolvement of the continuous state in each discrete state $q(t) = q$ is:
\begin{equation}
    \frac{{d{\mathbf{x}}\left( t \right)}}{{dt}} = {{\mathbf{b}}_q}.
    \label{eq5}
\end{equation}

Note that the evolution of AoS either increases with a slope of 1 or remains unchanged at value 0. So in order to describe AoS, the continuous state vector ${\mathbf{x}}\left( t \right)$ in SHS requires only one dimension ${x_0}\left( t \right)$, corresponding ${{\mathbf{b}}_q}=b_q$, where $b_q$ is a binary value. $b_q=1$ corresponds to the unit rate growth of ${x_0}\left( t \right)$ in discrete state $q$, and $b_q=0$ corresponds to ${x_0}\left( t \right)$ keeping constant in state $q$. Similarly, the values of ${{\mathbf{A}}_l}={A}_l$ are only 0 and 1, ${A}_l=0$ means that the update is completed and the AoS is reduced to 0, and ${A}_l=1$ means that the AoS keeps increasing.

At the same time, when using SHS method to calculate the average energy consumption and average AoS, it is necessary to calculate the stationary state probability of Markov chain and the correlation vector between discrete state ${q}\left( t \right)$ and continuous state ${x_0}\left( t \right)$. Let ${{\mathbf{\pi}} _q}\left( t \right)$ denote the probability that the Markov chain is in state $q$ , and ${{\mathbf{v}}_q}\left( t \right) = {v_{{q_0}}}\left( t \right)$ denotes the correlation between the discrete state and the continuous state. Therefore, we can obtain
\begin{equation}
    {{\mathbf{\pi}} _q}\left( t \right) = \Pr \left( {q\left( t \right) = q} \right) = {\rm E}\left[ {{\delta _{q,q\left( t \right)}}} \right].
    \label{eq6}
\end{equation}
\begin{equation}
    {{\mathbf{v}}_q}\left( t \right) = {v_{{q_0}}}\left( t \right) = {\rm E}\left[ {{x_0}\left( t \right){\delta _{q,q\left( t \right)}}} \right].
    \label{eq7}
\end{equation}

Let $L_q$ represent the set of all transitions at state $q$, and $L'_q$ denote the set of transitions passed in at state $q$. One of the basic assumptions for this kind of analysis is that Markov chain $q(t)$ is ergodic. Under this assumption, the state probability vector ${\mathbf{\pi }}\left( t \right) = \left[ {{\pi _0}\left( t \right) \ldots {\pi _m}\left( t \right)} \right]$ always converges to the only constant vector ${\mathbf{\bar \pi }} = \left[ {{{\bar \pi }_0} \ldots {{\bar \pi }_m}} \right]$ which satisfies
\begin{subequations}\label{eq8}
\begin{equation}
    {\bar \pi _q}\sum\nolimits_{l \in {L_q}} {{\lambda ^{\left( l \right)}}}  = \sum\nolimits_{l \in {L'_q}} {{\lambda ^{\left( l \right)}}} {\bar \pi _{{q_l}}},\quad q \in \mathbb{Q},
    \label{eq8a}
\end{equation}
\begin{equation}
    \sum\nolimits_{q \in \mathbb{Q}} {{{\bar \pi }_q} = 1} ,
    \label{eq8b}
\end{equation}
\end{subequations}

In addition, if the Markov chain of the discrete state is ergodic and stationary according to ${\mathbf{\bar \pi }}$, it has been shown in \cite[Theorem 4]{SHS for AoI} that there must be a non-negative solution such that
\begin{equation}
    {{\mathbf{\bar v}}_q}\sum\nolimits_{l \in {L_q}} {{\lambda ^{\left( l \right)}}}  = {{\mathbf{b}}_q}{\bar \pi _q} + \sum\nolimits_{l \in {{L'}_q}} {{\lambda ^{\left( l \right)}}} {{\mathbf{\bar v}}_{{q_l}}}{{\mathbf{A}}_l},\quad q \in \mathbb{Q},
    \label{eq9}
\end{equation}
then the average AoS is given by
\begin{equation}
    \bar\Delta  = \sum\nolimits_{q \in \mathbb{Q}} {{v_{{q_0}}}}.
    \label{eq10}
\end{equation}
And the average energy consumption can be obtained by
\begin{equation}
    {\rm E}\left[ P \right] = \sum\limits_{q \in \mathbb{Q}} {{{\bar \pi }_q}{P_q}} ,\quad {P_q} \in \left\{ {{P_{\textrm{B}}},{P_{\textrm{I}}},{P_{\textrm{S}}},{P_{\textrm{W}}}} \right\}.
    \label{eq11}
\end{equation}

\subsection{Analysis with SHS}
In this part, we derive the average AoS and average energy consumption for different wake-up policies with SHS.

\subsubsection{N-policy}

In N-policy, the discrete state space of the Markov chain is $\mathbb{Q} = \left\{ {B,ID,SL,1,2, \cdots ,N} \right\}$. In particular, $B$ refers to busy state, $ID$ refers to idle state, $SL$ refers to sleep state, and $k \in \{1, 2, \cdots, N\}$ refers to the state that a total of $k$ packets have arrived during the sleep state. Note that $q\left(t\right)=N$ equivalently represents wake-up state, as the server immediately turns to this state when the $N^{th}$ packet arrives. The continuous state degrades to a scalar $x_0\left(t\right) = \Delta \left(t\right)$, which is the AoS of the source.

\begin{figure}[htb]
\centering
\includegraphics[width=40mm]{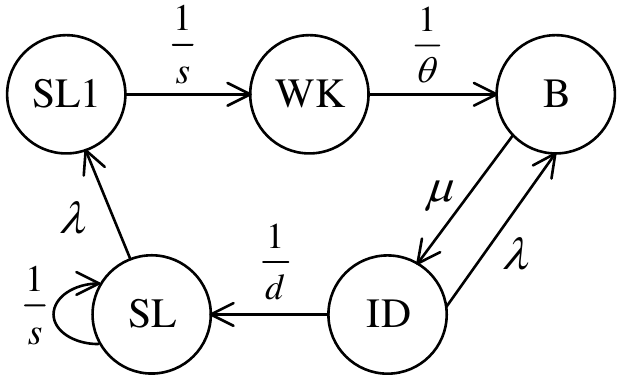}
\caption{State transition of N-policy.}
\label{Single source non-ideal N-policy}
\end{figure}

The state transitions for the discrete state $q\left( t \right)$ is shown in Fig. \ref{Single source non-ideal N-policy}. The corresponding transitions of continuous state ${\mathbf{x}}\left( t \right)$ are summarized in the Table \ref{table:Single source non-ideal N-policy}. The transitions are detailed as follows.

\begin{table}[htbp]
\renewcommand\arraystretch{1.2}
\begin{center}
\caption{Table of Transitions for the Markov Chain in Fig. \ref{Single source non-ideal N-policy}.}
\label{table:Single source non-ideal N-policy}
\begin{tabular}{ | c | c | c | c | c | c | }
\hline
     $l$   & ${{q_l} \to {{q'}_l}}$   & ${{\lambda ^{\left( l \right)}}}$   & ${{\mathbf{x}}{{\mathbf{A}}_l}}$   & ${{{\mathbf{A}}_l}}$   & ${{{{\mathbf{\bar v}}}_{{q_l}}}{{\mathbf{A}}_l}}$    \\
\hline
     $1$   & ${B \to ID}$   & $\mu$    & ${\left[ 0 \right]}$   & ${\left[ 0 \right]}$   & ${\left[ 0 \right]} $   \\
\hline
      $2$   & ${ID \to B}$   & $\lambda$    & ${\left[ {{x_0}} \right]}$   & ${\left[ 1 \right]}$   & ${\left[ {{v_{ID0}}} \right]}$   \\
\hline
      $3$   & ${ID \to SL}$   & ${\frac{1}{d}}$   & ${\left[ {{x_0}} \right]}$   & ${\left[ 1 \right]}$   & ${\left[ {{v_{ID0}}} \right]} $   \\
\hline
      $4$   & ${SL \to 1}$   & $\lambda$    & ${\left[ {{x_0}} \right]}$   & ${\left[ 1 \right]}$   & ${\left[ {{v_{SL0}}} \right]}$   \\
\hline
      $5$   & ${1 \to 2}$   & $\lambda$    & ${\left[ {{x_0}} \right]}$   & ${\left[ 1 \right]}$   & ${\left[ {{v_{10}}} \right]}  $  \\
\hline
      $ \cdots $   & $ \cdots $   &$ \cdots $   &$ \cdots $   &$ \cdots $   &$ \cdots $        \\
\hline
      $N+3$   & ${N-1 \to N}$   & $\lambda$    & ${\left[ {{x_0}} \right]}$   & ${\left[ 1 \right]}$   & ${\left[ {{v_{20}}} \right]}$     \\
\hline
     $N+4$   & ${N \to B}$   & ${\frac{1}{\theta }}$   & ${\left[ {{x_0}} \right]}$   & ${\left[ 1 \right]}$   & ${\left[ {{v_{N0}}} \right]}$   \\
\hline
\end{tabular}
\end{center}
\end{table}

\begin{itemize}
\item[-] \textit{$l=1$}: When a packet completes service and is delivered to the monitor, the server state changes from busy to idle. In this transition, the data on the monitor side is synchronized with source. Thus, the AoS of source becomes zero, i.e., $x'_0 = 0$.
\item[-] \textit{$l=2$}: When a packet arrives in the idle state, the server turns to busy state. In this case, the AoS of source remains the same, that is, $x'_0 = x_0$, because the arrival of the packet does not reduce the AoS until it is completely processed.
\item[-] \textit{$l=3$}: If no packets arrive during the idle state, the server turns to sleep state. In this transition, $x'_0=x_0$, because no packets are processed.
\item[-] \textit{$l=4,5, \cdots, N+3$}: The state turns from state $l-4$ to $l-3$ when a new packet arrives in the sleep state, where state 0 is equivalent to state $SL$. This transition also does not change $x_0$.
\item[-] \textit{$l=N+4$}: When a total of $N$ packets arrive during sleep state, the server immediately turns to wake-up state, and then turns to busy state after a while. This transition also does not change $x_0$.
\end{itemize}

The evolution of ${\mathbf{x}}\left( t \right)$ is determined by the discrete state $q\left( t \right)$. Specifically, when $q\left( t \right)=q$, we have
\begin{equation}
    \frac{{d{\mathbf{x}}\left( t \right)}}{{dt}} = {{\mathbf{b}}_q} = \left\{ \begin{gathered}
  \left[ 1 \right],q \in \{1, 2, \cdots, N, B\} \hfill \\
  \left[ 0 \right],q \in \{ID,SL\} \hfill \\
\end{gathered}  \right.
\label{eq.bq_Single source non-ideal N-policy}
\end{equation}
The explanation for (\ref{eq.bq_Single source non-ideal N-policy}) is that $q \in \{1, 2, \cdots, N, B\}$ means that there are unprocessed packets in the system. Thus, the AoS grows at a unit rate. $q \in \{ID,SL\}$ means that there are no packets in the system. Hence, the source and the monitor are synchronized and the AoS remains constant at 0.

With the above conditions, we can calculate the average energy consumption and average AoS by solving ${{{\bar \pi }_q}}$ and ${{\mathbf{\bar v}}_q}$. Firstly, we use (\ref{eq8a}) and (\ref{eq8b}) to calculate the stationary probability vector ${\mathbf{\bar \pi } = \left[{\bar\pi _B},{\bar\pi _{ID}},{\bar\pi _{SL}},{\bar\pi _1},{\bar\pi _2}, \cdots ,{\bar\pi _N}\right]}$. The matrix form of (\ref{eq8a}) can be expressed as ${\mathbf{\bar \pi D}} = {\mathbf{\bar \pi Q}}$, where $\mathbf{D}$ and $\mathbf{Q}$ are given as \[{\mathbf{D}} = diag\left[ {\mu ,\lambda  + \frac{1}{d},\underbrace {\lambda ,\lambda , \cdots ,\lambda }_{N elements },\frac{1}{\theta }} \right],\] \[{\mathbf{Q}} = \left[ {\begin{array}{*{20}{c}}
  0&\mu &0&0&0&{}&0 \\
  \lambda &0&{\frac{1}{d}}&0&0&{}&0 \\
  0&0&0&\lambda &0& \vdots &0 \\
  0&0&0&0&\lambda &{}&0 \\
  {}&{}& \cdots &{}&{}& \ddots &{} \\
  0&0&0&0&0&{}&\lambda  \\
  {\frac{1}{\theta }}&0&0&0&0& \cdots &0
\end{array}} \right].\]
With ${\mathbf{\bar \pi D}} = {\mathbf{\bar \pi Q}}$ and $\sum\nolimits_{q \in \mathbb{Q}} {{{\bar \pi }_q} = 1}$, we can obtain the stationary probability of each state as
\begin{subequations}\label{eq:pi_Single source non-ideal N-policy}
\begin{equation}
    {{\bar \pi }_B} = A \cdot \frac{{1 + d\lambda }}{\mu },
\end{equation}
\begin{equation}
    {{\bar \pi }_{ID}} = A \cdot d,
\end{equation}
\begin{equation}
    {{\bar \pi }_{SL}} = {{\bar \pi }_1} =  \cdots  = {{\bar \pi }_{N - 1}} = A \cdot \frac{1}{\lambda },
\end{equation}
\begin{equation}
    {{\bar \pi }_N} = A \cdot \theta,
\end{equation}
\end{subequations}
where $A = \frac{1}{{\frac{N}{\lambda } + \frac{1}{\mu } + \theta  + d\left( {1 + \frac{\lambda }{\mu }} \right)}}$.

By substituting (\ref{eq:pi_Single source non-ideal N-policy}) into (\ref{eq9}), the value of ${\mathbf{\bar v}}_q$ is obtained. Further substituting ${\mathbf{\bar v}}_q$ into (\ref{eq10}), the average AoS is given as
\begin{equation}
    \bar \Delta  = \frac{{\frac{{1 + d\lambda }}{\mu } + \frac{\theta }{\mu } + {\theta ^2} + \frac{{N\left( {N - 1} \right)}}{{2{\lambda ^2}}} + \frac{{N - 1}}{\lambda }\left( {\theta  + \frac{1}{\mu }} \right)}}{{\frac{N}{\lambda } + \frac{1}{\mu } + \theta  + d\left( {1 + \frac{\lambda }{\mu }} \right)}}.
\end{equation}
And the average energy consumption can be obtained by substituting (\ref{eq:pi_Single source non-ideal N-policy}) into (\ref{eq11}), which is expressed as
\begin{equation}
  {\text{E}}\left[ P \right] = \frac{{\frac{{1 + d\lambda }}{\mu }{P_{\textrm{B}}} {+} d{P_{\textrm{I}}} {+} \frac{N}{\lambda }{P_{\textrm{S}}} {+} \theta {P_{\textrm{W}}}}}{{\frac{N}{\lambda } + \frac{1}{\mu } + \theta  + d\left( {1 + \frac{\lambda }{\mu }} \right)}}.
\end{equation}

\subsubsection{Single-sleep policy} \label{Single-sleep policy}

In single-sleep policy, the discrete state space of the Markov chain is $\mathbb{Q} = \left\{ {SL,SL1,WK,WK1,B,ID0,ID} \right\}$. To distinguish whether the server is busy when a new packet arrives, we define two sets of states. Specifically, the states $SL$, $WK$, $ID0$, $ID$ indicate that the server is in sleep state, wake-up state, idle state after wake-up and idle state after processing, respectively, without packets in the system. Note that the difference between $ID0$ and $ID$ is that the server will turn to $SL$ if no packets arrive during $ID$ state, but not in $ID0$. The states $SL1$, $WK1$, $B$, indicate that the server is in sleep state, wake-up state, busy state, respectively, all with packets in the queue or in processing. The continuous state also degrades to a scalar $x_0\left(t\right) = \Delta \left(t\right)$.

\begin{figure}[htb]
\centering
\includegraphics[width=50mm]{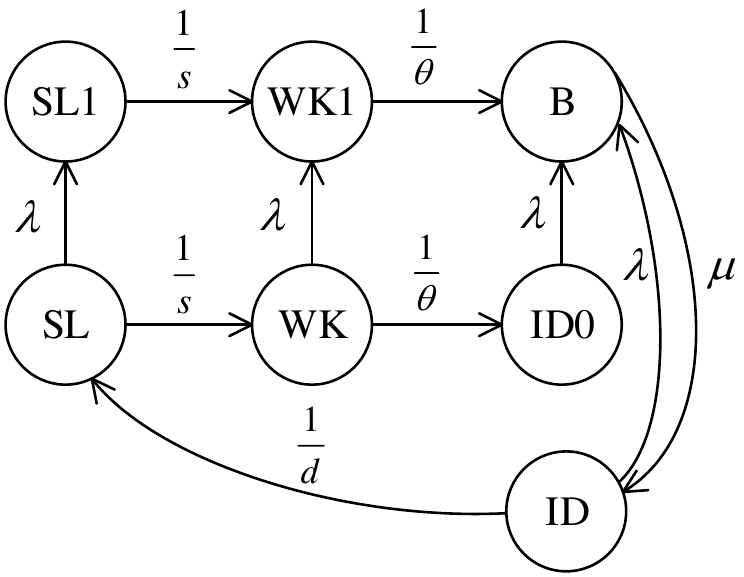}
\caption{State transition of single-sleep policy for a single source.}
\label{Single source non-ideal single-sleep}
\end{figure}

The state transition for the discrete state $q\left( t \right)$ is shown in Fig. \ref{Single source non-ideal single-sleep}. The corresponding transitions of continuous state ${\mathbf{x}}\left( t \right)$ are summarized in Table \ref{table:Single source non-ideal single-sleep}. The transitions are detailed as follows.

\begin{table}[htbp]
\renewcommand\arraystretch{1.2}
\begin{center}
\caption{Table of Transitions for the Markov Chain in Fig.\ref{Single source non-ideal single-sleep}.}
\label{table:Single source non-ideal single-sleep}
\begin{tabular}{ | c | c | c | c | c | c | }
\hline
     $l$   & ${{q_l} \to {{q'}_l}}$   & ${{\lambda ^{\left( l \right)}}}$   & ${{\mathbf{x}}{{\mathbf{A}}_l}}$   & ${{{\mathbf{A}}_l}}$   & ${{{{\mathbf{\bar v}}}_{{q_l}}}{{\mathbf{A}}_l}}$    \\
\hline
     $1$   & ${SL \to SL1}$   & $\lambda$    & ${\left[ {{x_0}} \right]}$   & ${\left[ 1 \right]}$   & ${\left[ {{v_{SL0}}} \right]} $   \\
\hline
      $2$   & ${SL \to WK}$   & ${\frac{1}{s}}$    & ${\left[ {{x_0}} \right]}$   & ${\left[ 1 \right]}$   & ${\left[ {{v_{SL0}}} \right]}$   \\
\hline
      $3$   & ${SL1 \to WK1}$   & ${\frac{1}{s}}$   & ${\left[ {{x_0}} \right]}$   & ${\left[ 1 \right]}$   & ${\left[ {{v_{SL10}}} \right]} $   \\
\hline
      $4$   & ${WK \to WK1}$   & $\lambda$    & ${\left[ {{x_0}} \right]}$   & ${\left[ 1 \right]}$   & ${\left[ {{v_{WK0}}} \right]}  $  \\
\hline
      $5$   & ${WK \to ID0}$   & ${\frac{1}{\theta }}$    & ${\left[ {{x_0}} \right]}$   & ${\left[ 1 \right]}$   & ${\left[ {{v_{WK0}}} \right]}$   \\
\hline
      $6$   & ${WK1 \to B}$   & ${\frac{1}{\theta }}$    & ${\left[ {{x_0}} \right]}$   & ${\left[ 1 \right]}$   & ${\left[ {{v_{WK10}}} \right]}$   \\
\hline
      $7$   & ${ID0 \to B}$   & $\lambda$    & ${\left[ {{x_0}} \right]}$   & ${\left[ 1 \right]}$   & ${\left[ {{v_{ID0}}} \right]}$   \\
\hline
      $8$   & $ {B \to ID} $   &$ \mu $   &$ {\left[ 0 \right]} $   &$ {\left[ 0 \right]} $   &$ {\left[ 0 \right]} $        \\
\hline
      $9$   & ${ID \to B}$   & $\lambda$    & ${\left[ {{x_0}} \right]}$   & ${\left[ 1 \right]}$   & ${\left[ {{v_{ID0}}} \right]}$     \\
\hline
     $10$   & ${ID \to SL}$   & ${\frac{1}{d }}$   & ${\left[ {{x_0}} \right]}$   & ${\left[ 1 \right]}$   & ${\left[ {{v_{ID0}}} \right]}$   \\
\hline
\end{tabular}
\end{center}
\end{table}

\begin{itemize}
\item[-] \textit{$l=1,4$}: Packets arrive when the server is in the sleep or wake-up state. In these transitions, the server turns into state SL1 or WK1. The AoS remains the same, that is, $x'_0 = x_0$, because the arrival of the packet does not reduce the AoS until it is completely processed.
\item[-] \textit{$l=2,3$}: After the server has been in sleep state for a period of time, it turns into wake-up state. With this transition, the AoS of source does not change, that is, $x'_0 = x_0$.
\item[-] \textit{$l=5$}: If no packets arrive during WK state, it turns into ID0 state to keep idling and wait for a packet arrival. With this transition, $x'_0 = x_0$.
\item[-] \textit{$l=6$}: After the server has been in WK1 state for a period of time, it turns into busy state. With this transition, $x'_0 = x_0$.
\item[-] \textit{$l=7$}: Packets arrive when the server is in ID0 state. In this transition, the server turns into busy state. With this transition, $x'_0 = x_0$.
\item[-] \textit{$l=8$}: A packet completes service, the server state changes from busy to idle. In this transition, $x'_0 = 0$.
\item[-] \textit{$l=9,10$}: The same as $l=2,3$ in Table \ref{table:Single source non-ideal N-policy}.
\end{itemize}

The evolution of ${\mathbf{x}}\left( t \right)$ is determined by the discrete state $q\left( t \right)$. Specifically, when $q\left( t \right)=q$, we have
\begin{equation}
    \frac{{d{\mathbf{x}}\left( t \right)}}{{dt}} = {{\mathbf{b}}_q} = \left\{ \begin{gathered}
  \left[ 1 \right],q \in \{SL1, WK1, B\} \hfill \\
  \left[ 0 \right],q \in \{SL, WK, ID0, ID\} \hfill \\
\end{gathered}  \right.
\label{eq.bq_Single source non-ideal single-policy}
\end{equation}
The explanation for (\ref{eq.bq_Single source non-ideal single-policy}) is that $q \in \{SL1, WK1, B\}$ means that there are unprocessed packets in the system. Thus, the AoS grows at a unit rate. $q \in \{SL, WK, ID0, ID\}$ means that there are no packets in the system. Hence, the source and the monitor are synchronized and the AoS at 0.

% With the above conditions, we can calculate the average energy consumption and average AoS by solving ${{{\bar \pi }_q}}$ and ${{\mathbf{\bar v}}_q}$. Firstly, we use (\ref{eq8a}) and (\ref{eq8b}) to calculate the stationary probability vector ${\mathbf{\bar \pi } = \left[{\bar\pi _{SL}},{\bar\pi _{SL1}},{\bar\pi _{WK}},{\bar\pi _{WK1}},{\bar\pi _{B}},{\bar\pi _{B1}},{\bar\pi _{ID}}\right]}$. By substituting $\mathbf{\bar \pi }$ into (\ref{eq9}), the value of ${\mathbf{\bar v}}_q$ is obtained.

% Further substituting ${\mathbf{\bar v}}_q$ into (\ref{eq10}) and $\mathbf{\bar \pi }$  into (\ref{eq11}), the average AoS and average energy consumption are given as
With the above conditions, we can calculate the average energy consumption and average AoS by solving ${{{\bar \pi }_q}}$ and ${{\mathbf{\bar v}}_q}$. Firstly, we use (\ref{eq8a}) and (\ref{eq8b}) to calculate the stationary probability vector ${\mathbf{\bar \pi } = \left[{\bar\pi _{SL}},{\bar\pi _{SL1}},{\bar\pi _{WK}},{\bar\pi _{WK1}},{\bar\pi _{B}},{\bar\pi _{B1}},{\bar\pi _{ID}}\right]}$. The matrix form of (\ref{eq8a}) can be expressed as ${\mathbf{\bar \pi D}} = {\mathbf{\bar \pi Q}}$, where $\mathbf{D}$ and $\mathbf{Q}$ are given as \[{\mathbf{D}} = diag\left[ {\lambda  + \frac{1}{s} ,\frac{1}{s},\lambda  + \frac{1}{\theta} ,\frac{1}{\theta} ,\lambda,\mu,\lambda+  \frac{1}{d }} \right],\] \[{\mathbf{Q}} = \left[ {\begin{array}{*{20}{c}}
  0&\lambda &\frac{1}{s}&0&0&0&0 \\
  0&0&0&{\frac{1}{d}}&0&0&0 \\
  0&0&0&\lambda &{\frac{1}{\theta}}&0&0 \\
  0&0&0&0&0&{\frac{1}{\theta}}&0 \\
  0&0&0&0&0& \lambda &0 \\
  0&0&0&0&0&0&\mu  \\
  {\frac{1}{d}}&0&0&0&0&  \lambda &0
\end{array}} \right].\]
With ${\mathbf{\bar \pi D}} = {\mathbf{\bar \pi Q}}$ and $\sum\nolimits_{q \in \mathbb{Q}} {{{\bar \pi }_q} = 1}$, we can obtain the stationary probability of each state as
\begin{subequations}\label{eq:pi_Single source non-ideal single-sleep}
\begin{equation}
    {\bar \pi _{SL}} = \frac{1}{B}\left[ {s\mu \lambda \left( {\theta \lambda  + 1} \right)} \right],
\end{equation}
\begin{equation}
    {\bar \pi _{SL1}} = \frac{1}{B}\left[ {{s^2}\mu {\lambda ^2}\left( {\theta \lambda  + 1} \right)} \right],
\end{equation}
\begin{equation}
    {\bar \pi _{WK}} = \frac{1}{B}\theta \mu \lambda,
\end{equation}
\begin{equation}
    {\bar \pi _{WK1}} = \frac{1}{B}\left[ {\theta \mu {\lambda ^2}\left( {s + \theta  + s\theta \lambda } \right)} \right],
\end{equation}
\begin{equation}
    {\bar \pi _{ID0}} = \frac{1}{B}\mu ,
\end{equation}
\begin{equation}
    {\bar \pi _B} = \frac{1}{B}\left[ {\lambda \left( {d\lambda  + 1} \right)\left( {s\lambda  + 1} \right)\left( {\theta \lambda  + 1} \right)} \right],
\end{equation}
\begin{equation}
    {\bar \pi _{ID}} = \frac{1}{B}\left[ {d\mu \lambda \left( {s\lambda  + 1} \right)\left( {\theta \lambda  + 1} \right)} \right],
\end{equation}
\end{subequations}
where \[\begin{gathered}
  B = \mu  + \lambda  + ds\theta {\lambda ^4} + \left( {d + s + \theta } \right)\left( {{\lambda ^2} + \mu \lambda  + s\theta \mu {\lambda ^3}} \right) \hfill \\
   + \left( {{s^2} + {\theta ^2} + 2s\theta  + ds + d\theta } \right)\mu {\lambda ^2} + \left( {ds + s\theta  + \theta d} \right){\lambda ^3}. \hfill \\
\end{gathered} \]

By substituting (\ref{eq:pi_Single source non-ideal single-sleep}) into (\ref{eq9}), the value of ${\mathbf{\bar v}}_q$ is obtained. Further substituting ${\mathbf{\bar v}}_q$ into (\ref{eq10}) and $\mathbf{\bar \pi }$  into (\ref{eq11}), the average AoS and average energy consumption are given as
\begin{equation}
    \bar \Delta   = \frac{{\lambda C}}{{\mu B}},
\end{equation}
\begin{equation}
  {\text{E}}\left[ P \right] = \frac{{s{P_{\textrm{S}}} {+} \theta {P_{\textrm{W}}} {+} ( {d {+} \frac{1}{{D}}} ){P_{\textrm{I}}} {+} \frac{{d\lambda  {+} 1}}{\mu }{P_{\textrm{B}}}}}{{\frac{B}{\mu D}}},
\end{equation}
where  \[\begin{gathered}
  C = \left( {{\mu ^2}{\lambda ^2}s\theta  {+} \mu \lambda } \right)\left( {{s^2} {+} s\theta  {+} {\theta ^2}} \right) {+} {\mu ^2}\lambda \left( {{s^3} {+} {s^2}\theta  {+} s{\theta ^2} {+} {\theta ^3}} \right) \hfill \\
   {+} \mu {\lambda ^2}s\theta \left( {s {+} \theta } \right) {+} {\lambda ^3}ds\theta  {+} {\lambda ^2}\left( {ds {+} s\theta  {+} \theta d} \right) {+} \lambda \left( {d {+} s {+} \theta } \right) {+} 1,
\end{gathered} \]
$D = \lambda \left( {s\lambda  {+} 1} \right)\left( {\theta \lambda  {+} 1} \right)$.

\subsubsection{Multi-sleep policy}

In Multi-sleep policy, the discrete state space of the Markov chain is $\mathbb{Q} = \left\{ {SL,SL1,WK,B,ID} \right\}$. In particular, all the states have the same meaning as the ones with the same name in \ref{Single-sleep policy}. The continuous state also degrades to a scalar $x_0\left(t\right) = \Delta \left(t\right)$.

\begin{figure}[htb]
\centering
\includegraphics[width=40mm]{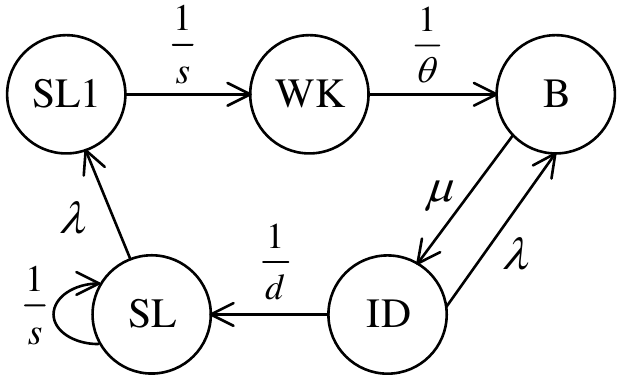}
\caption{State transition of multi-sleep policy for a single sources.}
\label{Single source non-ideal multi-sleep}
\end{figure}

The state transition for the discrete state $q\left( t \right)$ is shown in Fig. \ref{Single source non-ideal multi-sleep}. The corresponding transitions of continuous state ${\mathbf{x}}\left( t \right)$ are summarized in Table \ref{table:Single source non-ideal multi-sleep}, which is very similar to Table \ref{table:Single source non-ideal single-sleep}. $l=1,3,4,5,6,7$ is the same as $l=1,3,5,8,9,10$ in Table \ref{table:Single source non-ideal single-sleep}. The main difference is the self-transition $l = 2$, which means that the server sleeps again after one sleep period if there is no packets arrival during this period.

\begin{table}[htbp]
\renewcommand\arraystretch{1.2}
\begin{center}
\caption{Table of Transitions for the Markov Chain in Fig.\ref{Single source non-ideal multi-sleep}.}
\label{table:Single source non-ideal multi-sleep}
\begin{tabular}{ | c | c | c | c | c | c | }
\hline
     $l$   & ${{q_l} \to {{q'}_l}}$   & ${{\lambda ^{\left( l \right)}}}$   & ${{\mathbf{x}}{{\mathbf{A}}_l}}$   & ${{{\mathbf{A}}_l}}$   & ${{{{\mathbf{\bar v}}}_{{q_l}}}{{\mathbf{A}}_l}}$    \\
\hline
     $1$   & ${SL \to SL1}$   & $\lambda$    & ${\left[ {{x_0}} \right]}$   & ${\left[ 1 \right]}$   & ${\left[ {{v_{SL0}}} \right]} $   \\
\hline
      $2$   & ${SL \to SL}$   & ${\frac{1}{s}}$    & ${\left[ {{x_0}} \right]}$   & ${\left[ 1 \right]}$   & ${\left[ {{v_{SL0}}} \right]}$   \\
\hline
      $3$   & ${SL1 \to WK}$   & ${\frac{1}{s}}$   & ${\left[ {{x_0}} \right]}$   & ${\left[ 1 \right]}$   & ${\left[ {{v_{SL10}}} \right]} $   \\
\hline
      $4$   & ${WK \to B}$   & ${\frac{1}{\theta }}$    & ${\left[ {{x_0}} \right]}$   & ${\left[ 1 \right]}$   & ${\left[ {{v_{WK0}}} \right]}$   \\
\hline
      $5$   & $ {B \to ID} $   &$ \mu $   &$ {\left[ 0 \right]} $   &$ {\left[ 0 \right]} $   &$ {\left[ 0 \right]} $        \\
\hline
      $6$   & ${ID \to B}$   & $\lambda$    & ${\left[ {{x_0}} \right]}$   & ${\left[ 1 \right]}$   & ${\left[ {{v_{ID0}}} \right]}$     \\
\hline
      $7$   & ${ID \to SL}$   & ${\frac{1}{d }}$   & ${\left[ {{x_0}} \right]}$   & ${\left[ 1 \right]}$   & ${\left[ {{v_{ID0}}} \right]}$   \\
\hline
\end{tabular}
\end{center}
\end{table}

% Using method similar to that mentioned above, the average AoS and average energy consumption can be obtained as
The evolution of ${\mathbf{x}}\left( t \right)$ is determined by the discrete state $q\left( t \right)$. Specifically, when $q\left( t \right)=q$, we have
\begin{equation}
    \frac{{d{\mathbf{x}}\left( t \right)}}{{dt}} = {{\mathbf{b}}_q} = \left\{ \begin{gathered}
  \left[ 1 \right],q \in \{SL1,WK,B\} \hfill \\
  \left[ 0 \right],q \in \{SL,ID\} \hfill \\
\end{gathered}  \right.
\label{eq.bq_Single source non-ideal multi-sleep}
\end{equation}
The explanation for (\ref{eq.bq_Single source non-ideal multi-sleep}) is that $q \in \{SL1,WK,B\}$ means that there are unprocessed packets in the system. Thus, the AoS grows at a unit rate. $q \in \{SL,ID\}$ means there are no packets in the system. Hence, the source and the monitor are synchronized and the AoS remains 0.

With the above conditions, we can calculate the average energy consumption and average AoS by solving ${{{\bar \pi }_q}}$ and ${{\mathbf{\bar v}}_q}$. Firstly, we use (\ref{eq8a}) and (\ref{eq8b}) to calculate the stationary probability vector ${\mathbf{\bar \pi } = \left[{\bar\pi _{SL}},{\bar\pi _{SL1}},{\bar\pi _{WK}},{\bar\pi _{B}},{\bar\pi _{ID}}\right]}$. The matrix form of (\ref{eq8a}) can be expressed as ${\mathbf{\bar \pi D}} = {\mathbf{\bar \pi Q}}$, where $\mathbf{D}$ and $\mathbf{Q}$ are given as \[{\mathbf{D}} = diag\left[ {\lambda  + \frac{1}{s} ,\frac{1}{s}, \frac{1}{\theta},\mu,\lambda+  \frac{1}{d }} \right],\] \[{\mathbf{Q}} = \left[ {\begin{array}{*{20}{c}}
  \frac{1}{s}&\lambda &0&0&0 \\
  0&0&\frac{1}{s}&0&0 \\
  0&0&0&{\frac{1}{\theta}}&0 \\
  0&0&0&0&\mu \\
  {\frac{1}{d}}&0&0&  \lambda &0
\end{array}} \right].\]
With ${\mathbf{\bar \pi D}} = {\mathbf{\bar \pi Q}}$ and $\sum\nolimits_{q \in \mathbb{Q}} {{{\bar \pi }_q} = 1}$, we can obtain the stationary probability of each state as
\begin{subequations}\label{eq:pi_Single source non-ideal multi-sleep}
\begin{equation}
    {{\bar \pi }_{SL}} = E \cdot \mu,
\end{equation}
\begin{equation}
    {{\bar \pi }_{SL1}} = E \cdot s\mu\lambda,
\end{equation}
\begin{equation}
    {{\bar \pi }_WK} = E \cdot \theta\mu\lambda,
\end{equation}
\begin{equation}
    {{\bar \pi }_B} = E \cdot \lambda \left(1 + d\lambda \right),
\end{equation}
\begin{equation}
    {{\bar \pi }_{ID}} = E \cdot d\mu\lambda,
\end{equation}
\end{subequations}
where $E = \frac{1}{{\mu  + \lambda  + d{\lambda ^2} + d\mu \lambda  + s\mu \lambda  + \theta \mu \lambda }}$.

By substituting (\ref{eq:pi_Single source non-ideal multi-sleep}) into (\ref{eq9}), the value of ${\mathbf{\bar v}}_q$ is obtained. Further substituting ${\mathbf{\bar v}}_q$ into (\ref{eq10}) and $\mathbf{\bar \pi }$  into (\ref{eq11}), the average AoS and average energy consumption are given as
\begin{equation}
    \bar \Delta  = \frac{{\lambda \left( {{s^2}{\mu ^2} {+} s\theta {\mu ^2} {+} s\mu  {+} {\theta ^2}{\mu ^2} {+} \theta \mu  {+} d\lambda  {+} 1} \right)}}{{\mu \left( {\mu  {+} \lambda  {+} d{\lambda ^2} {+} d\mu \lambda  {+} s\mu \lambda  {+} \theta \mu \lambda } \right)}},
\end{equation}
\begin{equation}
  {\text{E}}\left[ P \right] = \frac{{\mu \left( {s\lambda  {+} 1} \right){P_{\textrm{S}}} {+} \theta \mu \lambda {P_{\textrm{W}}} {+} \lambda \left( {d\lambda  {+} 1} \right){P_{\textrm{B}}} {+} d\mu \lambda {P_{\textrm{I}}}}}{{\mu  {+} \lambda  {+} d{\lambda ^2} {+} d\mu \lambda  {+} s\mu \lambda  {+} \theta \mu \lambda }}.
\end{equation}

\section{Simulation Results}
In this section, we present the trade-off between average AoS and average energy consumption for the different policies through Monte Carlo simulations. In the following, we assume that the energy consumed in each state in the sleep model is ${P_{\textrm{B}}}=1$, ${P_{\textrm{I}}}={P_{\textrm{W}}}=0.5$, ${P_{\textrm{S}}}=0$, as per the assumptions made in our study.

\begin{figure}[htb]
\centering
\includegraphics[width=90mm]{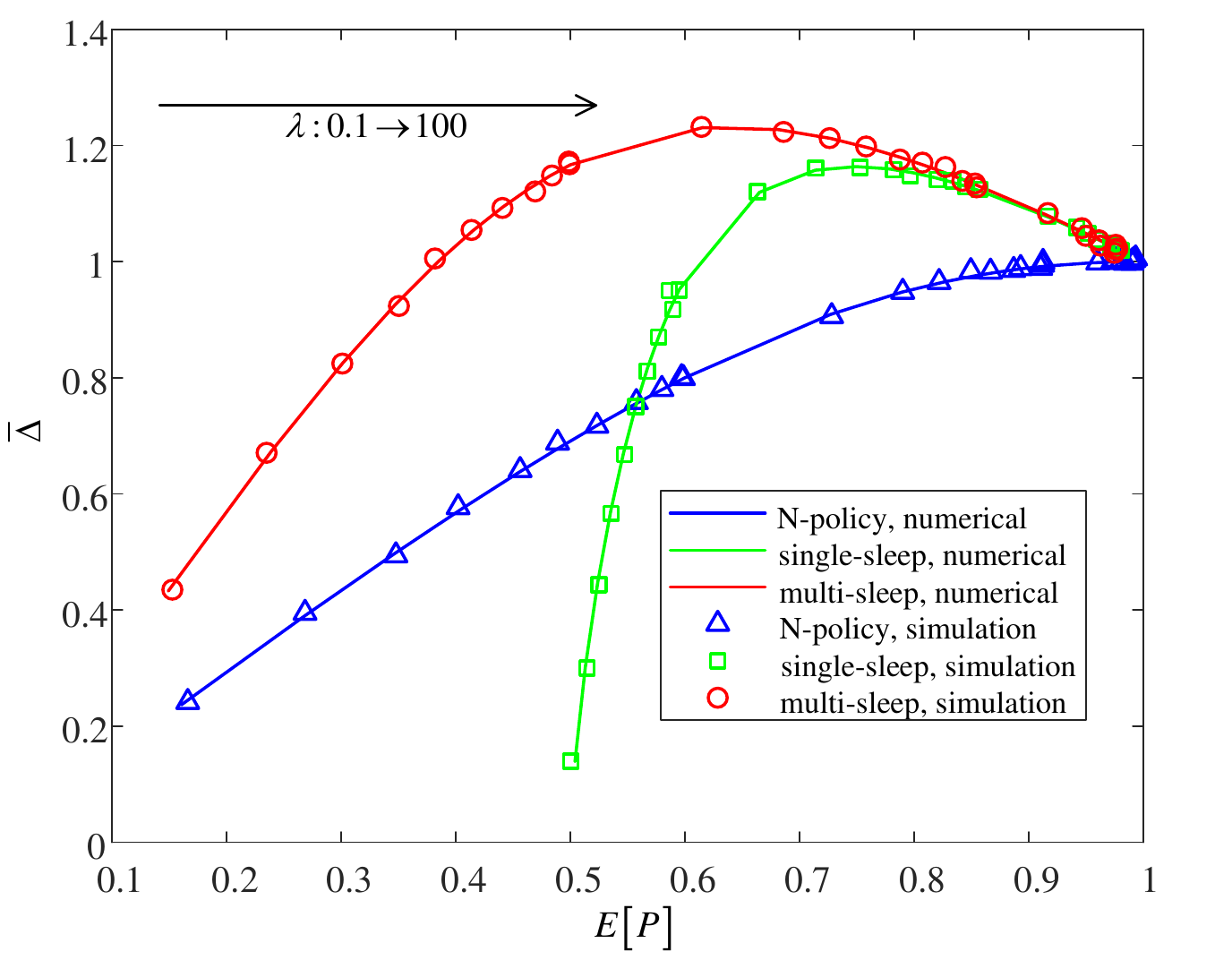}
\caption{Age-energy trade-off versus packets arrival rate $\lambda \in [0.1, 100]$, with
$\mu = 1$, $N = s =1$, $d = \theta = 1$.}
\label{fig_lambda}
\end{figure}
Fig. \ref{fig_lambda} depicts the trade-off between average AoS and average energy consumption under different packet arrival rates $\lambda$. The simulation results align well with the analytical ones, validating our theoretical analysis. It is observed that the average energy consumption increases with the increase of $\lambda$ in all the curves. This is due to the fact that the higher the $\lambda$, the less likely the system is in a sleep state and the more likely it is in a busy state, leading to larger energy consumption. However, for average AoS, it does not always increase with the increase of $\lambda$. The curves of single-sleep and multi-sleep exhibit a trend of increasing first and then decreasing. This is because the increase of $\lambda$ has a double effect on average AoS. When $\lambda$ is small, its increase results in less time for information to be synchronized, thereby increasing the average AoS. When $\lambda$ exceeds a certain threshold, its increase increases the probability that the server transitions directly from the idle state to the busy state, thereby accelerating the packet processing process, leading to a reduction in the average AoS. Moreover, when $\lambda$ is large and tends to infinity, the results of the three policies are indistinguishable. This is because the server does not enter sleep state but always transitions directly from the idle state to the busy state.

\begin{figure}[htb]
\centering
\includegraphics[width=90mm]{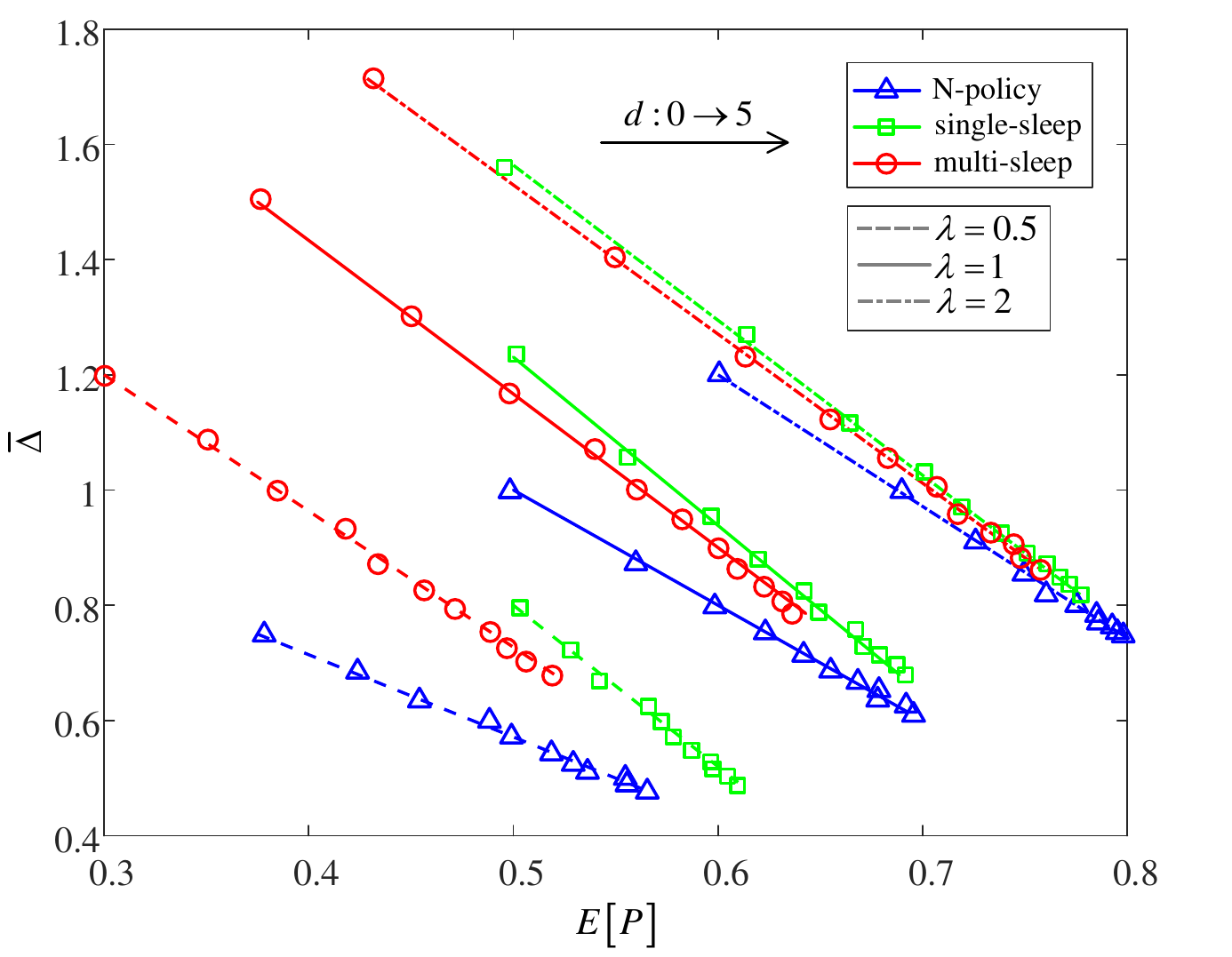}
\caption{The impact of $d$ on the age-energy trade-off  with $\mu = 1$, $N = s =1$, $\theta = 1$.}
\label{fig_one_source_d}
\end{figure}
Then, we analyze the trade-off between average AoS and average energy consumption under different system parameters in each wake-up policy. The impact of idling time $d$ on the trade-off curve is shown in Fig. \ref{fig_one_source_d}. As $d$ increases, the average energy consumption increases while the average AoS decreases. This is because a higher $d$ leads to less time in the sleep state and more time processing packets in the busy state.

\begin{figure}[htb]
\centering
\includegraphics[width=90mm]{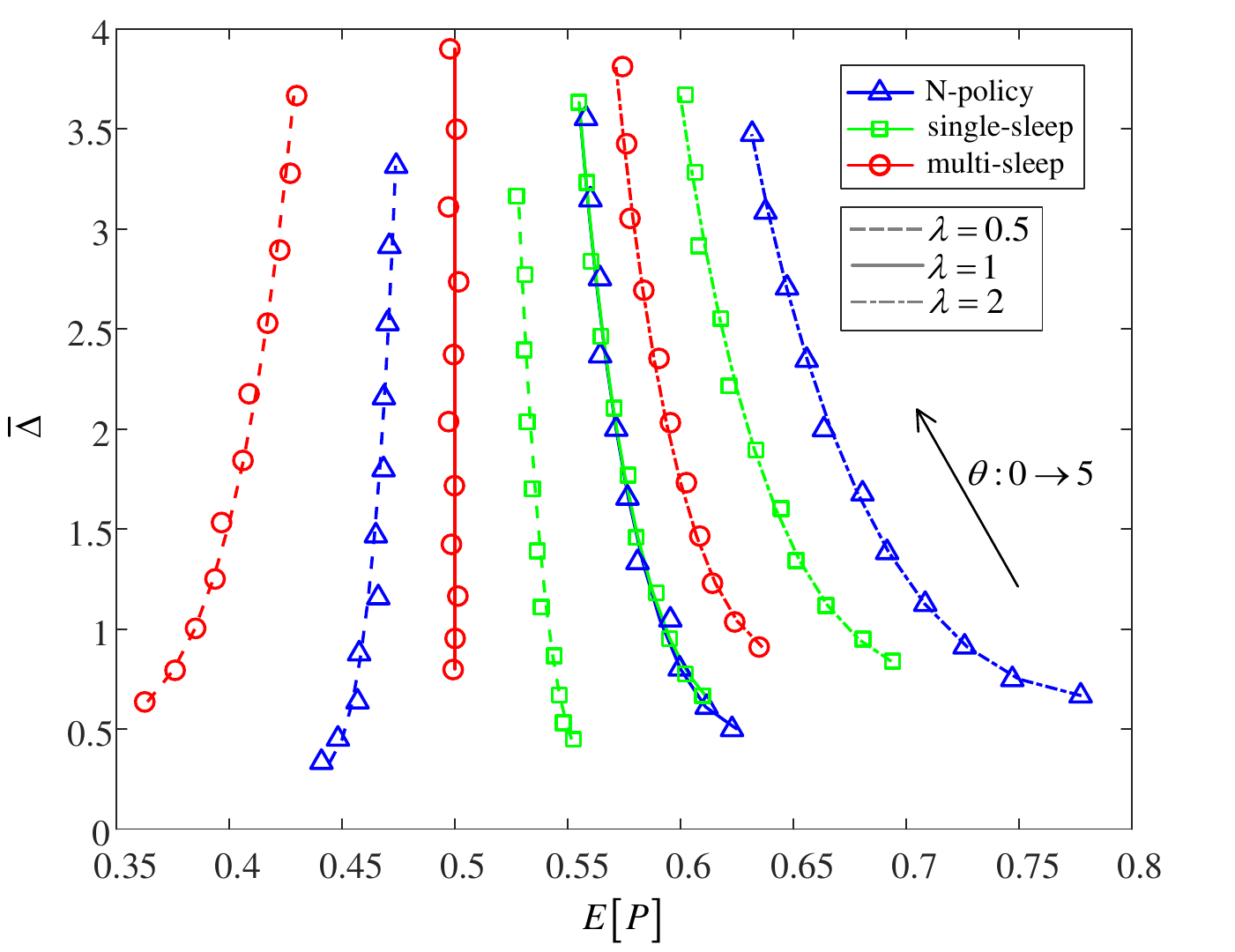}
\caption{The impact of $\theta$ on the age-energy trade-off with $\mu = 1$, $N = s =1$, $d=1$.}
\label{fig_one_source_theta}
\end{figure}
Fig. \ref{fig_one_source_theta} shows the impact of wake-up time $\theta$ on the trade-off curve. The average AoS increases as $\theta$ increases because it slows down the packet processing, resulting in a larger average AoS. The average energy consumption, however, shows different trends in different system loads. In the case of low system load, where the arrival rate $\lambda$ is less than the service rate of $\mu$, the server is in a low-power sleep state most of the time. Increasing $\theta$ increases the time in the wake-up state, resulting in an increase in average energy consumption. Conversely, in the case of heavy system load, where the server is busy most of the time, increasing the sleep time reduces the average energy consumption.

\begin{figure}[htb]
\centering
\includegraphics[width=90mm]{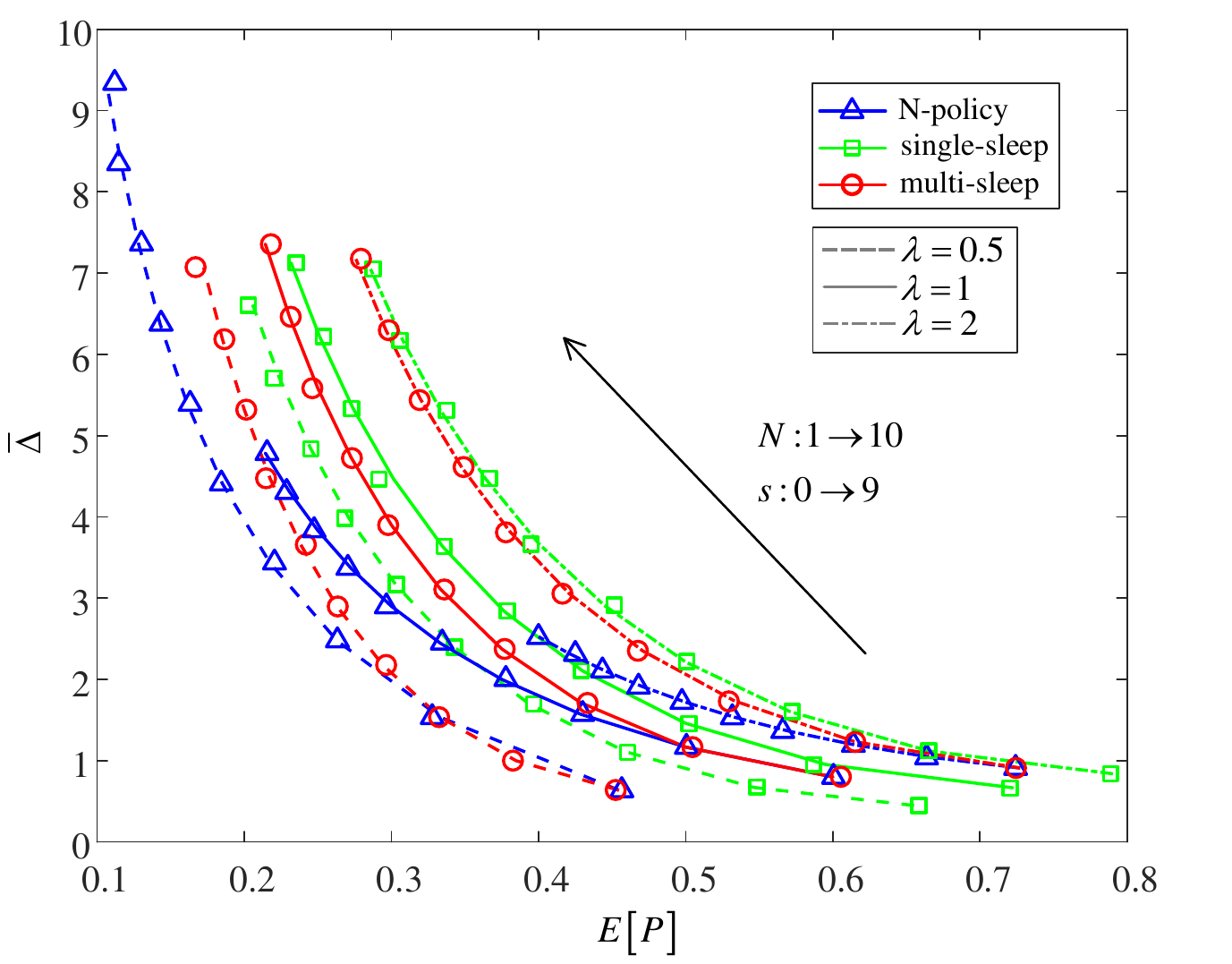}
\caption{The impact of $N$ and $s$ on the age-energy trade-off with $\mu = 1$, $d = \theta =1$.}
\label{fig_one_source}
\end{figure}
Fig. \ref{fig_one_source} illustrates the impact of sleep parameters ($N$ and $s$) on the age-energy trade-off curve for different wake-up policies. The results show that as $N$ or $s$ increases, the average energy consumption decreases, while the average AoS increases in all policies. This is due to the direct effect of increasing the sleeping time, which sacrifices the AoS to save energy. Comparison of the three wake-up policies reveals that N-policy performs best in terms of trade-off. Moreover, when the arrival rate of $\lambda$ is low, it is not suitable to adopt N-policy with a higher $N$ value because it makes it difficult to wake up when $N$ is large, leading to severe out-of-sync issues. Furthermore, it is observed that as the arrival rate of $\lambda$ increases, the trade-off curves of single-sleep policy and multi-sleep policy almost coincide because fast packet arrival makes it unlikely that there will be no packet arrival within a sleep duration $s$, and the effect of multi-sleep policy is almost equivalent to that of single-sleep policy.

\section{Conclusion}\label{IV}
In this paper, we investigate a status update system comprising a source and a server, and propose a sleep model to conserve energy when the server is idle. We introduce three wake-up policies to activate the server and derive explicit expressions of the average AoS and average energy consumption using the SHS method for each policy. We demonstrate the trade-offs between average AoS and average energy consumption under different system parameters via Monte Carlo simulation. Our simulation results show that, for a heavy system load, the sleep model effectively limits the growth of AoS. For a light system load, a small wake-up time $\theta$ can reduce both age and energy consumption. The trade-off scope that N-policy provides widens as the system load decreases. For a high system load, N-policy offers the best age performance, while single-sleep and multi-sleep policies tend to be comparable. In future work, we plan to explore the expansion of the system to multiple sources.

% \bibitem{1}
% C. Cho, C. Lin and J. Wang, ``Reliable grouping gaf algorithm using hexagonal virtual cell structure,'' in \textit{Proc. ICST'08}, Tainan, Nov. 2008.

\end{document}